# Improving Post-Processing for Quantitative Precipitation Forecasting Using Deep Learning: Learning Precipitation Physics from High-Resolution Observations

ChangJae Lee[a], Heecheol Yang[b], Byeonggwon Kim[a]

[a] *Korean Meteorological Administration, Seoul, Korea*

[b] *Chungnam National University, Daejeon, Korea.*

*Corresponding author*: Heecheol Yang, hcyang@cnu.ac.kr

# ABSTRACT

Accurate quantitative precipitation forecasting (QPF) remains one of the main challenges in numerical weather prediction (NWP), primarily due to the difficulty of representing the full complexity of atmospheric microphysics through parameterization schemes. This study introduces a deep learning-based post-processing model, DL-QPF, which diagnoses precipitation fields from meteorological forecasts by learning directly from high-resolution radar estimates precipitation.

The DL-QPF model is constructed using a Patch-conditional Generative Adversarial Network (Patch-cGAN) architecture combined with a U-Net generator and a discriminator. The generator learns meteorological features relevant to precipitation, while the adversarial loss from the discriminator encourages the generation of realistic rainfall patterns and distributions. Training is performed on three years of warm-season data over the Korean Peninsula, with input variables derived from ECMWF's Integrated Forecasting System High-Resolution forecast (IFS-HRES).

Model verification is conducted against multiple reference models, including global (IFS-HRES, KIM), regional (KIM-Regional, KIM-LENS), and AI-based (GraphCast) forecasts. Verification across multiple rainfall thresholds shows that DL-QPF achieves a frequency bias near one and superior success ratios. Particularly for heavy and intense rainfall events, DL-QPF outperforms both conventional NWP and an AI model, demonstrating improved skill in capturing high-intensity precipitation.

This study highlights the potential of observational data-driven deep learning approaches in post-processing QPF. By directly learning from observations, DL-QPF reduces systematic biases and enhances the realism of forecasted rainfall distributions. These results demonstrate the model's potential to enhance QPF realism and suggest broader applications in operational forecasting tasks such as extreme rainfall and lightning prediction.

# SIGNIFICANCE STATEMENT

Improving the accuracy of quantitative precipitation forecasts (QPF) is essential for weather-dependent decision-making, especially in the face of high-impact weather events. Conventional NWP models often suffer from biases in precipitation forecasts. This study demonstrates the potential of deep learning as a promising post-processing tool that can learn precipitation characteristics directly from high-quality observational data. By leveraging a Patch-cGAN architecture trained on radar-estimated rainfall, the proposed DL-QPF model produces realistic and de-biased precipitation fields at fine spatial scales. Benchmark results show that DL-QPF outperforms conventional and AI-based forecast models across multiple rainfall intensities, particularly for heavy rainfall events. These results underscore the value of data-driven deep-learning approaches for enhancing precipitation forecasts.

## 1. Introduction

Accurate precipitation forecasts are critical for mitigating flood risks, optimizing water resource management, and supporting disaster response. However, they remain one of the most challenging aspects of Numerical Weather Prediction (NWP). The primary source of this challenge lies in that it is too complex to replicate the full spectrum of atmospheric microphysics, so that microphysics is estimated through parametrization schemes in NWP models (Stensrud, 2007).

To capture nature's precipitation physics, we try to predict precipitation close to the high-resolution observation in this study. Precipitation can be viewed as a diagnostic variable (Larraondo et al., 2019), thus we use a deep-learning approach to find the underlying patterns between meteorological variables and observed precipitation. As this study focuses on establishing relationships between precipitation and other meteorological fields from NWP, our approach is applied as a post-processing technique for quantitative precipitation forecasts (QPF). We demonstrate the improvements in QPF results, emphasizing how closely they align with natural precipitation patterns, as evaluated across various metrics and case studies in the following sections.

The primary objectives of this study are the next three: (1) to demonstrate the potential of a deep learning approach that leverages the advantages of directly learning from high-quality observational data in predicting realistic precipitation, (2) to benchmark the QPF performance of the proposed model against conventional NWP models and an AI-based model, highlighting their respective strengths and limitations, and (3) to evaluate model performance through several representative case studies.

We are not the first to apply deep-learning methods to enhance QPF results. Prior works have been tried to diagnose precipitation using various deep-learning methods. For the global scale, GraphCast (Lam et al., 2023) and AIFS (Lang et al., 2024) successfully launched AI forecast models using the European Centre for Medium-Range Weather Forecasts (ECMWF) reanalysis fields (ERA5) (Hersbach et al., 2020), 0.25° × 0.25° resolution, and for the regional scale, more high-resolution estimated precipitation from the satellites and radars was used (Zhou et al., 2022).

In this study, we aim to generate precipitation fields that most accurately reflect natural rainfall patterns. To achieve this, we use high-resolution (500 m × 500 m) and high-accuracy radar-based precipitation estimates that capture fine-scale rainfall details and closely align with rain gauge observations. In addition to leveraging high-quality observational data, our proposed method adopts Generative Adversarial Networks (GAN, Goodfellow et al., 2014) to produce realistic and detailed rainfall distributions, leading to improved QPF performance, particularly in heavy rainfall events. Code is available at https://github.com/hunter3789/Deep-Learning-QPF.

## 2. Data and Methods

We trained an NWP post-processing model using three years of warm-season data, spanning from May to October for 2021 through 2023, sampled at three-hour intervals. This study assumes a perfect NWP forecast, following the Perfect Prognostic Method (PPM; Marzban et al., 2006), whereby each observed precipitation is paired with the recent corresponding forecast data. This approach can be extended at inference time to incorporate longer forecast lead times when applying the model for operational post-processing.

*2.1. Numerical Weather Prediction: Input Data*

To implement a post-processing model, we take ECMWF's Integrated Forecasting System High RESolution forecast (IFS-HRES; Rasp et al., 2020), a global forecast product with 0.1° × 0.1° latitude/longitude resolution, as input data. Atmospheric variables of 15 pressure levels and the surface level, along with the altitude data obtained from the Shuttle Radar Topography Mission (SRTM; Farr et al., 2007), described in Table 1, are used here. For training, we used recent forecast data from 6h to 15h forecast lead times, considering a spin-up period to ensure stabilized NWP results (Ma et al., 2021), from the 00 and 12 UTC runs.

| Surface variables | Atmospheric variables | Pressure Level (hPa): 15 levels |
|---|---|---|
| 2m temperature | U-component wind | 1000, 950, 925, 900, 850, 800, 700, 600, 500, 400, 300, 250, 200, 150, 100 |
| 2m dew-point temperature | V-component wind | |
| Mean sea level pressure | Temperature | |
| Surface net solar radiation | Relative humidity | |
| Altitude | Geopotential height | |

*Table 1. Input variables and levels for the proposed model*

*2.2. Radar estimated precipitation: Data for target value (labeling)*

For the observational data, we used radar-estimated precipitation data processed with the CLEANER algorithm (Oh et al., 2020) from the Korea Meteorological Administration (KMA). This dataset has a spatial resolution of 500 m × 500 m and a temporal frequency of 10 minutes. 3-h accumulated precipitation is used as labeled data in this experiment.

Figure 2 illustrates that radar-estimated precipitation closely aligns with rain gauge observations, which directly measure rainfall at the surface, compared to the reanalysis precipitation field (ERA5). The inconsistency between reanalysis fields and in-situ observations may arise from the low resolution of reanalysis data and its inherent limitations in convective parameterization. Figure 1 shows the locations of each radar site and its coverage, as well as the locations of the rain gauges. Since our proposed model uses radar-estimated precipitation as labels, we can expect the results to better capture the true nature of precipitation compared to models trained on reanalysis fields.

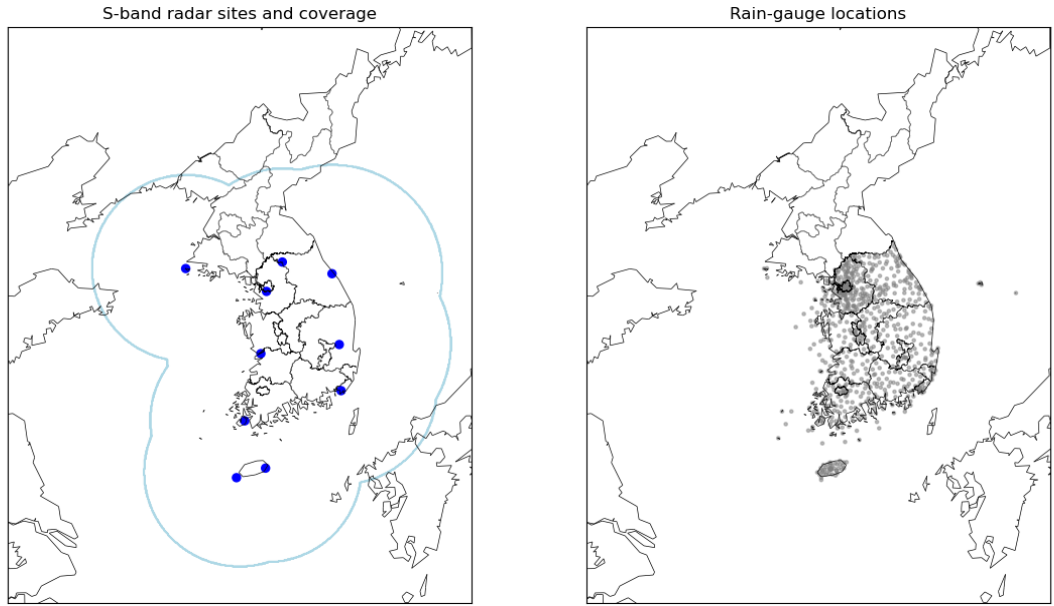

*Figure 1. Locations of S-band radar sites (blue dots) and their coverage areas (skyblue outline) (left), and locations (gray dots) of rain gauges (right) operated by the Korea Meteorological Administration (KMA).*

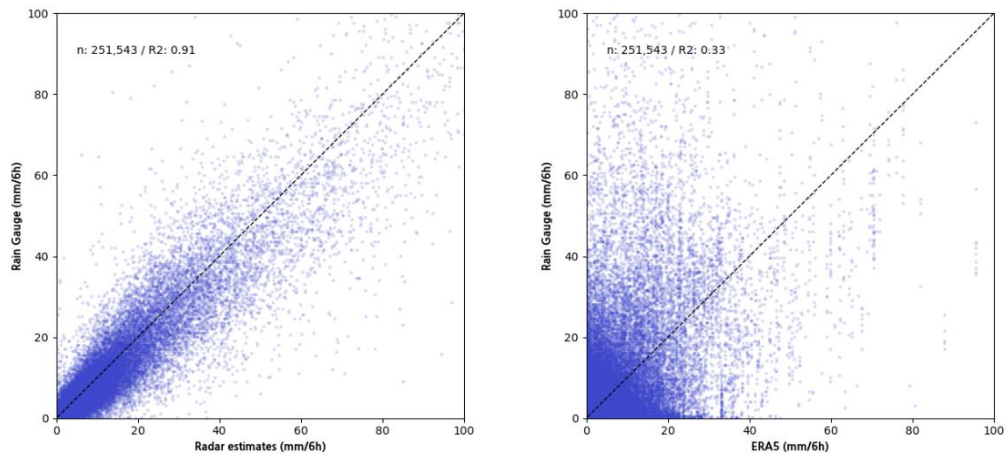

*Figure 2. Scatter plot comparing radar-estimated precipitation with observed rainfall from rain gauges (left), and comparison between ERA5-predicted precipitation fields and observed rainfall (right). Six-hour accumulated rainfall data from June to August 2024 is used. The comparison is based on the nearest-point method.*

*2.3. Pre-processing input and labeled data*

We designed our post-processing model to generate high-resolution precipitation fields conditioned on the input fields. However, due to the difference in spatial resolution and domain between the input and labeled data, we re-gridded them to a 2 km $\times$ 2 km resolution using the bilinear interpolation method and restricted the domain to around the Korean Peninsula region (1440 km $\times$ 1152 km).

After the re-gridding step, we normalized (standardized) our input and labeled data to neural networks to be trained efficiently. The labeled data is log-transformed before being normalized, since precipitation data has a highly-skewed distribution.

*2.4. Methods: Architecture*

The architecture of the proposed model is based on Patch-cGAN (Conditional Generative Adversarial Network; Isola et al., 2018), comprising a U-Net-based generator (Ronneberger et al., 2015) and a basic Convolutional Neural Network (CNN)-based discriminator.

The generator in our proposed model takes meteorological variables as input and produces corresponding precipitation fields. The generator's U-Net architecture captures both local and global features of input data through several consecutive convolutional layers with different sizes of kernels and residual connections, enabling it to predict precipitation at each grid of the input conditioned on meteorological variables. We leveraged a simple U-Net architecture here and replaced the max-pooling layers with convolutional layers, which have learnable parameters, to preserve contextual information within the kernel better. Further details are provided in Figure 3. The input data have a horizontal dimension of 720 $\times$ 576, corresponding to a 1440 km $\times$ 1152 km domain, and include 80 meteorological variables.

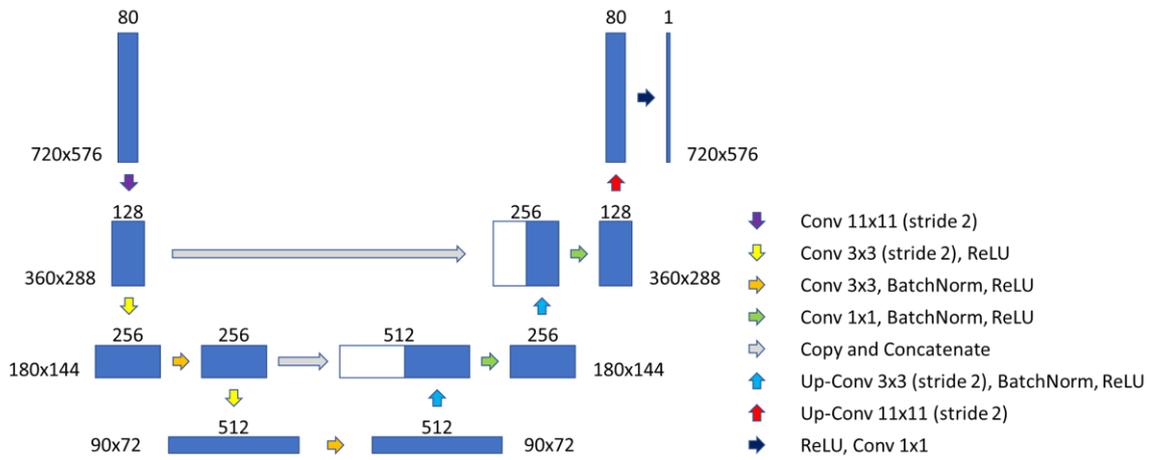

*Figure 3. Architecture of the generator. Each blue box represents a feature map, with the number of latent dimensions indicated above the box and the horizontal size labeled alongside. White boxes denote copied feature maps passed through residual connections. Color-coded arrows indicate different operations such as convolution, up-sampling, or concatenation.*

To enhance the model's ability to learn the natural distribution of precipitation, we implemented a discriminator that distinguishes between observed precipitation (the ground truth) and generated data. This discriminator learns to classify labeled data as real and generated data from the generator as fake for each patch, conditioned on the input of the generator. Each patch has a receptive field size of $135 \times 135$ pixels, corresponding to a 270 km $\times$ 270 km mesoscale size region. Details are described in Figure 4. The discriminator plays a key role in the Patch-cGAN architecture, with its loss functioning as the GAN loss that flows through the generator, encouraging it to produce sharp and realistic precipitation fields.

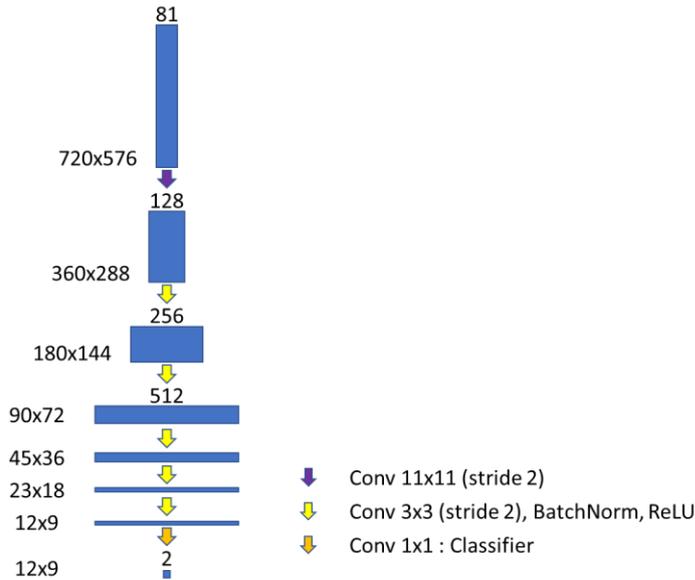

*Figure 4. Architecture of the discriminator. Each blue box represents a feature map, with the number of latent dimensions indicated above the box and the horizontal size labeled alongside. Color-coded arrows denote different operations such as convolution or activation functions.*

The total loss combines L1 loss (Mean Absolute Error) and L2 loss (Mean Squared Error) to ensure that the predicted values maintain an appropriate scale, with a GAN loss component, implemented as a cross-entropy loss. The loss function is defined as:

$$\mathcal{L}(G, D) = \mathbb{E}[logD(x,y)] + \mathbb{E}[\log(1 - D(x, G(x)))] + \lambda\{\alpha\mathcal{L}_{L1}(G) + (1-\alpha)\mathcal{L}_{L2}(G)\}, \quad (1)$$

where G denotes the generator mapping input variables x to output y, D denotes the discriminator, $\lambda = 100$, and $\alpha = 0.8$. This combined loss encourages the generator to produce realistic precipitation fields that can effectively fool the discriminator. This model is trained over 50 epochs, with a learning rate 10e-5.

Note that radar estimated precipitation (labeled data) is masked (Figure 1), as it is obtained from 10 S-band radar sites, each with a 240 km cover range. This masking needs to be considered when calculating the loss.

The primary goal of the proposed model is to generate precipitation fields with a histogram closely matching observation. Achieving this objective leads to improving the accuracy of QPF, as long as the input, NWP predictions, provides precisely predicted atmospheric fields.

*2.5. Verification strategies*

A reliable QPF model must accurately predict precipitation across all intensity categories. To assess this capability, we benchmarked our proposed model against conventional NWP models, described in the next paragraph, and an AI-based model, GraphCast, across four precipitation intensity categories: light rain (≥1 mm/6h), moderate rain (≥10 mm/6h), heavy rain (≥25 mm/6h), and intense rain (≥50 mm/6h). Table 2 provides a summary of each model used in the verification.

|  | Type | Resolution | Boundary condition | Training data | Input data |
|---|---|---|---|---|---|
| IFS-HRES | Global model forecast | $0.1° \times 0.1°$ | - | - | - |
| Korean Integrated Model (KIM) | Global model forecast | 12km × 12km | - | - | - |
| KIM-Regional | Regional model forecast | 3km × 3km | KIM | - | - |
| KIM-Localized Ensemble (KIM-LENS) | Regional model forecast, Spatial Aligned Mean | 3km × 3km | KIM-Global Ensemble | - | - |
| GraphCast | AI-based model | $0.25° \times 0.25°$ | - | ERA5 (1979-2019) Fine-tuned on IFS-HRES (2016-2021) | IFS-HRES |
| DL-QPF | DL-based post-processing | 2km × 2km | - | IFS-HRES (2021-2023) | IFS-HRES |

*Table 2. Summary of the models used for benchmarking.*

Since precipitation performance is highly dependent on model resolution, both global and regional NWP models are included in this verification. For the global models, we used

ECMWF's IFS-HRES and the Korean Integrated Model (KIM, Hong et al., 2018), a global model with a 12 km × 12 km resolution. To compare with a high-resolution model that has advantages in predicting convective rainfall, we also included the KIM-Regional model, which has a 3 km × 3 km resolution over a domain covering the Korean Peninsula and is initialized from KIM.

Additionally, the KIM-Localized Ensemble (KIM-LENS), which shares the same resolution and domain as the KIM-Regional model and is initialized from the KIM Global Ensemble with 13 members (expanded to 25 members using the time-lagged method), is also included in the verification. Instead of using individual forecasts from each ensemble member, we employed the Spatial Aligned Mean (Lee et al., 2024) for evaluation. This method preserves the sharpness of precipitation patterns by accounting for spatial variability across members, addressing the smoothing effects of arithmetic ensemble means.

We also included GraphCast's precipitation forecast, which has a 0.25°×0.25° resolution and is initialized from IFS-HRES, to assess the performance of an AI-based model.

Performance diagrams (Roebber, 2009) are used to benchmark the performance of each model across various metrics (Wilks, D. S., 1995), including Probability of Detection (POD, equivalent to Sensitivity), Success Ratio (SR, equivalent to Precision), Critical Success Index (CSI, equivalent to Intersection over Union), and Frequency Bias (FB, equivalent to Prediction Bias). These diagrams allow for a comprehensive assessment, and we analyzed each model's characteristics and strengths across these metrics.

## 3. Verification results

We benchmarked the QPF performance of various models over the summer months (June, July, and August) of 2024, using forecast lead times ranging from 6 to 72 hours from the 00 UTC runs. For a fair comparison, all models' predicted rainfall fields were re-gridded to a 2 km × 2 km resolution prior to verification. Verification was conducted using radar-estimated precipitation over the Korean Peninsula region.

In the performance diagrams, we aggregated the metrics for forecast days 1, 2, and 3 to minimize the influence of diurnal cycle effects, allowing for a more straightforward interpretation. For reference, each forecast day is marked by its corresponding number in the charts.

For convenience, our proposed deep-learning-based QPF model will be referred to as DL-QPF from here on.

*3.1. Light rain (Figure 5)*

For the light rain threshold (1 mm/ 6 h), low-resolution global models, including GraphCast, tend to represent an over-forecasting bias, whereas high-resolution models, including the proposed QPF model, tend to show an under-forecasting tendency. IFS-HRES and DL-QPF are the two models closest to a frequency bias of 1, which is the ideal target.

Impressively, DL-QPF achieves the highest success ratio among all models. This result indicates that our proposed model effectively leverages the advantages gained from learning directly from observational data.

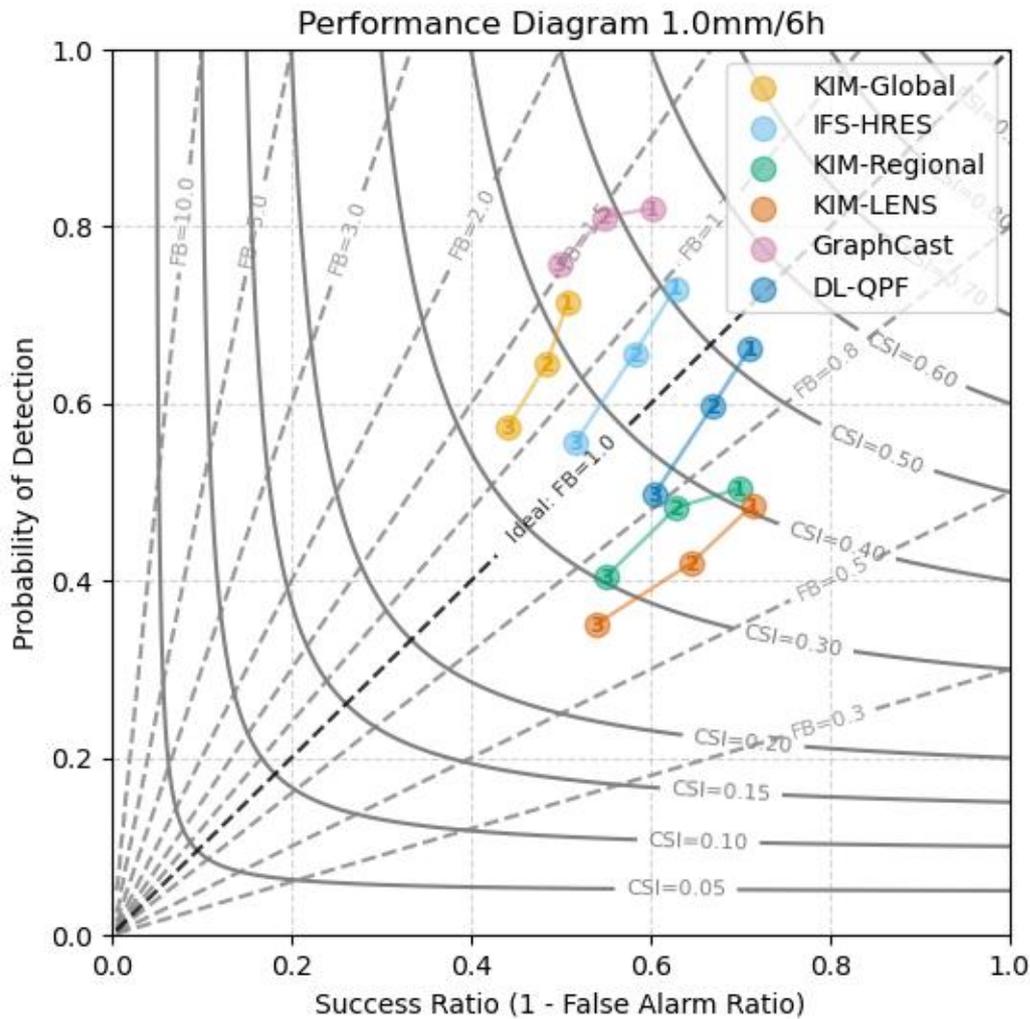

*Figure 5. Performance diagram for each model at the 1 mm/6 h rainfall threshold, from day 1 to day 3 forecasts during the summer of 2024. Each forecast day is indicated by its corresponding number in the chart.*

*3.2. Moderate rain (Figure 6)*

The moderate rain threshold (10 mm/ 6 h) is where the performance of all models tends to converge. All models except KIM-LENS predict with a frequency bias close to 1, while KIM-LENS shows an under-forecasting bias. Interestingly, GraphCast demonstrated strong performance at this threshold, despite being trained on the reanalysis precipitation fields rather than direct observations. DL-QPF ranked following GraphCast, showing better performance than conventional NWP models.

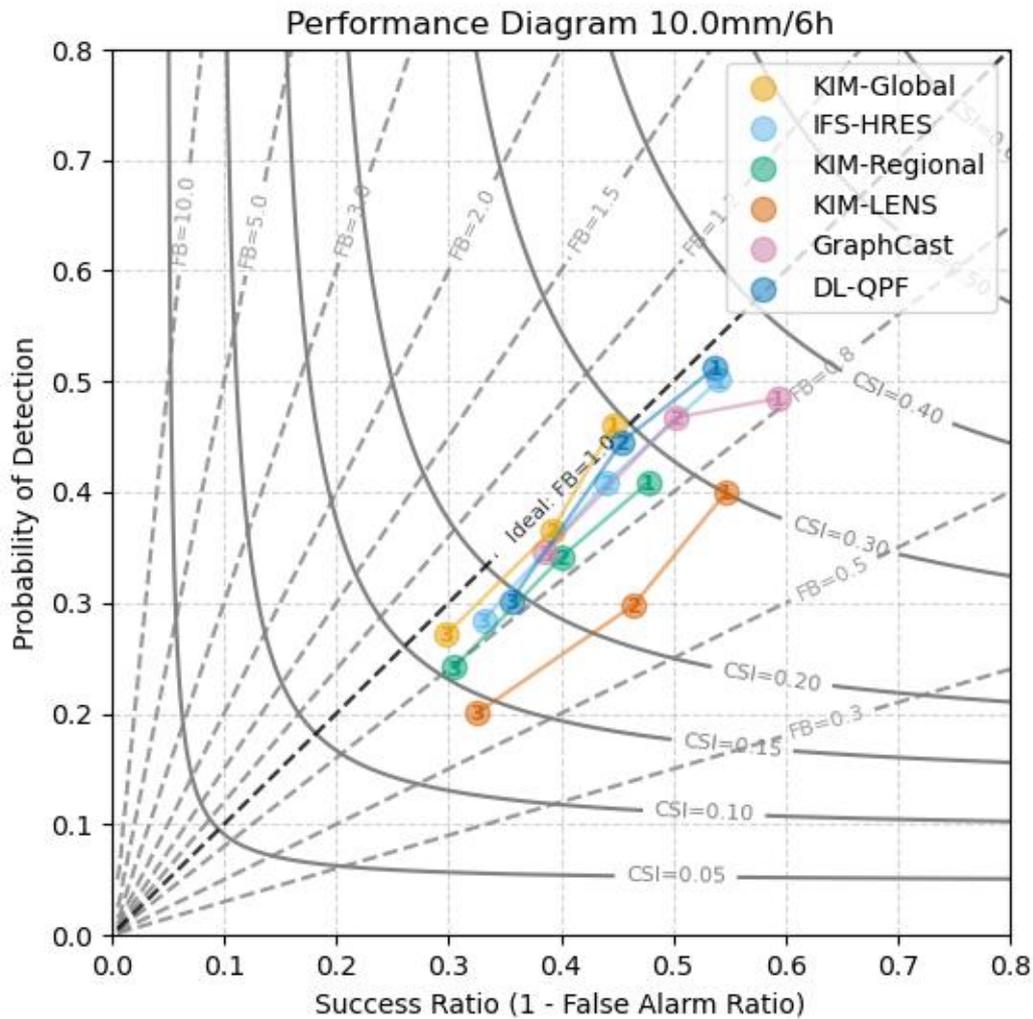

*Figure 6. Performance diagram for each model at the 10 mm/6 h rainfall threshold, from day 1 to day 3 forecasts during the summer of 2024. Each forecast day is indicated by its corresponding number in the chart.*

*3.3. Heavy rain (Figure 7)*

From the heavy rain threshold (25 mm/ 6 h), global models start to reveal their inherent weaknesses, primarily due to their low resolution and reliance on convective parameterization. Global models, including GraphCast, show a tendency toward under-forecasting heavy rain events, whereas high-resolution models, including the proposed DL-QPF model, predict with a frequency bias close to 1.

DL-QPF demonstrates the best CSI scores among all models, achieving good performances across all other metrics as well, with the highest POD, high SR, and a frequency bias closest to the ideal value of 1.

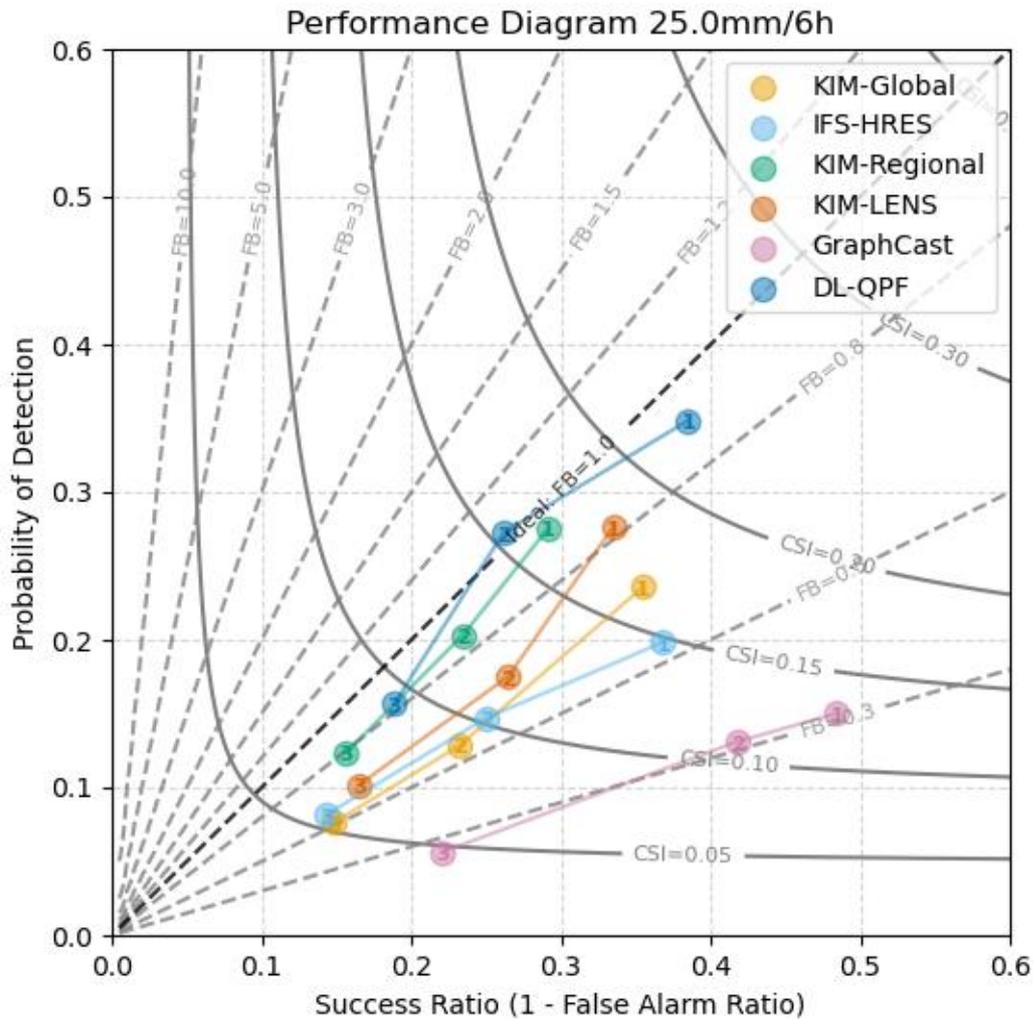

*Figure 7. Performance diagram for each model at the 25 mm/6 h rainfall threshold, from day 1 to day 3 forecasts during the summer of 2024. Each forecast day is indicated by its corresponding number in the chart.*

*3.4. Intense rain (Figure 8)*

For the intense rain threshold (50 mm/ 6 h), the same tendencies observed in the heavy rain category persist but with overall worse scores. At this threshold, GraphCast performed the worst among the models. This result highlights a well-known limitation of AI-based models: they tend to underestimate extreme events due to the nature of commonly used loss functions, such as L1 and L2 loss (Rasp et al., 2023).

However, DL-QPF achieved the best performance among all models, despite showing a slight under-forecasting tendency on day 1 forecasts. This performance can be attributed to the use of adversarial loss, which helps recover the true distribution of precipitation and mitigates the excessive smoothness typically caused by L1- or L2-based loss functions. High-resolution models, such as KIM-Regional and KIM-LENS, also performed relatively well at this threshold.

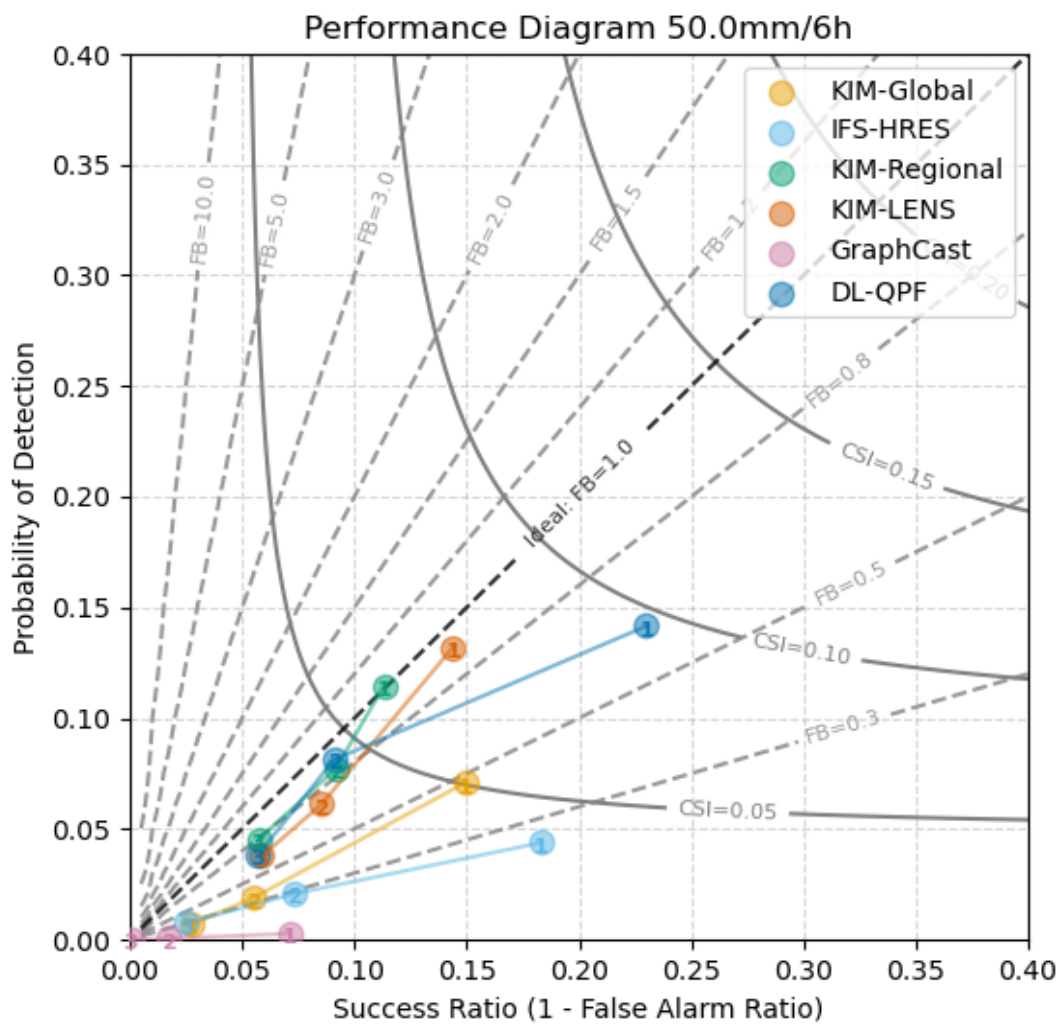

*Figure 8. Performance diagram for each model at the 50 mm/6 h rainfall threshold, from day 1 to day 3 forecasts during the summer of 2024. Each forecast day is indicated by its corresponding number in the chart.*

*3.5. Overall performance*

To summarize the results, our proposed model, DL-QPF, exhibits two main strengths: (1) its precipitation distribution closely matches observations, maintaining a frequency bias near 1 across all categories, and (2) it achieves a higher success ratio on average compared to other models. These characteristics contribute to overall performance, with particular improvements in the heavy rain and intense rainfall categories. This outcome highlights the advantage of our model learning precipitation features directly from observational data.

## 4. Case studies

In this section, we examined four cases to demonstrate how our proposed model performed in real-world scenarios, including a synoptic low, a stationary front with a meso-low, a convective shower event, and a tropical depression. We compared the outcomes, 3-h accumulated rainfall, of our model against radar observations and used the IFS-HRES predicted precipitation as a reference for evaluation.

We overlaid each precipitation field with the surface Mean Sea-Level Pressure (MSLP) field for better interpretation. Both IFS-HRES and DL-QPF share the same MSLP field, as DL-QPF uses the meteorological fields from IFS-HRES as input. The MSLP field composited on the radar observations is taken from the most recent forecast data to represent conditions as close as possible to the analysis field.

*4.1. Synoptic-Low case*

Figure 9 illustrates an intense rainfall event associated with a low-pressure system on 9 July 2024, which recorded more than 200 mm of rain over the day. In the predicted field, the low-pressure center is displaced slightly northward compared to the observation, causing the main precipitation region to also shift northward. Nevertheless, the overall distribution of precipitation predicted by DL-QPF more closely resembles the observed rainfall compared to that of IFS-HRES. Since our proposed model assumes a perfect forecast (PPM), the output is tied to the predicted meteorological fields, which explains the positional difference in the main precipitation region.

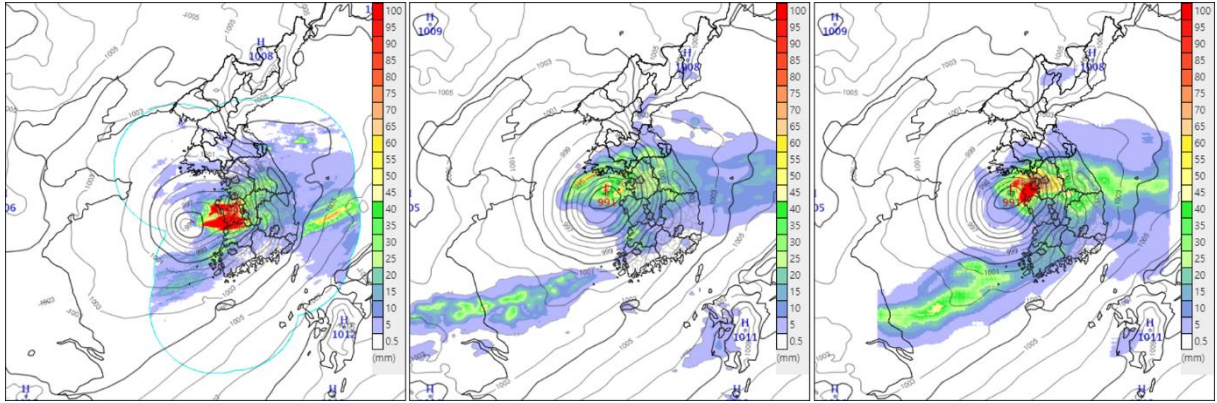

*Figure 9. Observed radar-estimated precipitation (left), IFS-HRES forecast (middle), and DL-QPF prediction (right) for 3-h accumulated precipitation associated with a synoptic low case, from the 42-h forecast valid at 18 UTC on 9 July 2024. Mean sea level pressure composites are taken from 6-h IFS-HRES forecast for the observation panel (left), and 42-h IFS-HRES forecast for the prediction panels (middle and right).*

*4.2. Stationary front with a meso-low case*

Figure 10 represents an event associated with a stationary front and a meso-low on 17 July 2024. Both the IFS-HRES and DL-QPF results show a broader precipitation region, whereas the observed precipitation is concentrated in a narrow band, which is a characteristic of a stationary front. DL-QPF better captures the intense rainfall areas and the overall distribution of precipitation intensity, although the observed rainfall exhibits a more compact and densely focused area with a greater total amount.

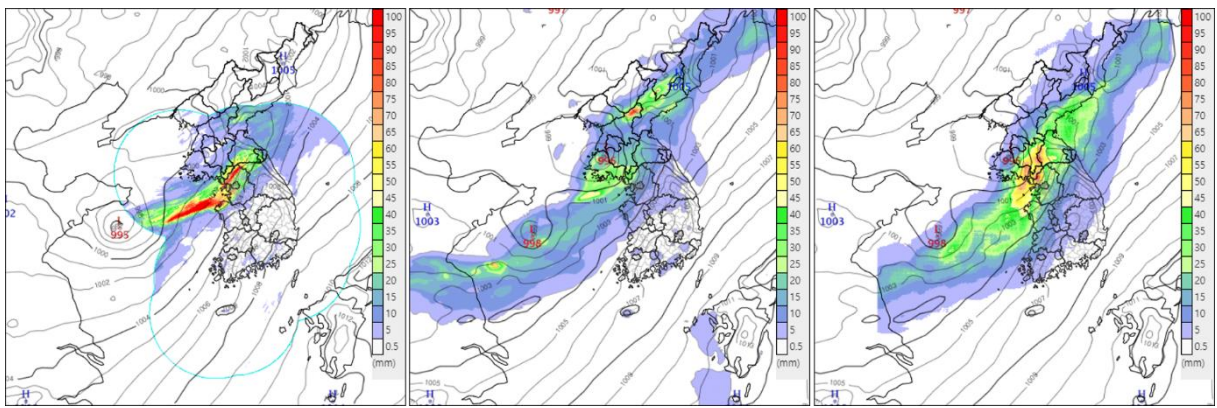

*Figure 10. Observed radar-estimated precipitation (left), IFS-HRES forecast (middle), and DL-QPF prediction (right) for 3-h accumulated precipitation associated with a stationary front case, from the 21-h forecast valid at 21 UTC on 17 July 2024. Mean sea level pressure composites are taken from 9-h IFS-HRES forecast for the observation panel (left), and 21-h IFS-HRES forecast for the prediction panels (middle and right).*

### 4.3. Convective shower case

Figure 11 depicts a convective shower event that occurred over a broad inland area of South Korea on 5 August 2024. DL-QPF shows a weakness in this case, underestimating both the extent and intensity of the convective precipitation. In contrast, IFS-HRES, aided by its convective parameterization scheme, better captures the potential regions of convective showers, although it also underestimates the precipitation intensity. Our proposed model exhibits lower confidence in handling convective shower cases during validation. This limitation is commonly observed even in advanced deep-learning models, which often struggle with uncertainty and confidence when predicting small-scale convective features (Ravuri et al., 2021).

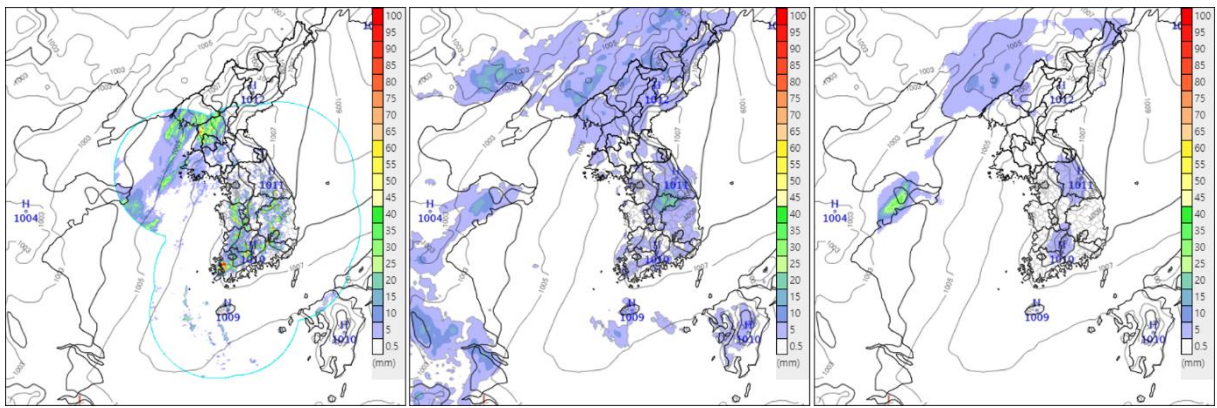

*Figure 11. Observed radar-estimated precipitation (left), IFS-HRES forecast (middle), and DL-QPF prediction (right) for 3-h accumulated precipitation associated with a convective shower case, from the 9-h forecast valid at 9 UTC on 5 August 2024. Mean sea level pressure composites are taken from 9-h IFS-HRES forecast.*

### 4.4. Tropical Depression (TD) case

Figure 12 displays a tropical depression (TD) case on 21 August 2024, which weakened from Typhoon Jongdari, the ninth typhoon of the 2024 season. As the typhoon moved into the Yellow Sea, it weakened to a TD and released moisture, enhancing heavy rainfall along the west coast of South Korea. Both IFS-HRES and DL-QPF captured the focused rainfall area near the TD. However, DL-QPF showed a better result than IFS-HRES in predicting precipitation outside the immediate TD region by avoiding over-forecasting, particularly in the Yellow Sea area south of the TD.

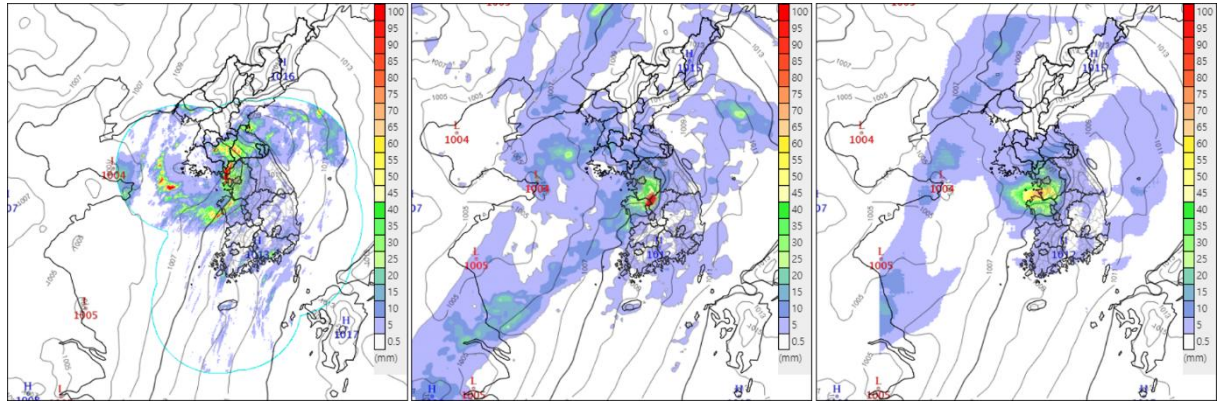

*Figure 12. Observed radar-estimated precipitation (left), IFS-HRES forecast (middle), and DL-QPF prediction (right) for 3-h accumulated precipitation associated with a TD case, from the 24-h forecast valid at 00 UTC on 21 August 2024. Mean sea level pressure composites are taken from 0-h IFS-HRES forecast for the observation panel (left), and 24-h IFS-HRES forecast for the prediction panels (middle and right).*

## 5. Summary and Discussion

From the benchmark results and case studies, we demonstrated that our proposed model, DL-QPF, effectively leverages the advantages of learning directly from observations by (1) producing a de-biased precipitation distribution that closely matches observations, and (2) achieving a higher success ratio compared to other models. These strengths contribute to the overall improvement of QPF results, although the model exhibited weakness in underestimating convective shower events due to the weaker relationship between meteorological fields and localized precipitation in such cases. This result implies that even with a simple architecture, QPF performance can be improved by learning directly from real-world observational data.

In this study, we intentionally did not use the NWP-predicted precipitation field as an input in order to avoid dependence on it and fulfill the objective of de-biasing its inherent bias.

However, as observed from our model's weakness in handling convective shower cases, incorporating the NWP precipitation field as an additional input may help in identifying potential convective rainfall areas through the guidance of the parameterization scheme, even though it may still struggle in accurately predicting precipitation intensity.

Since our model diagnoses the precipitation field at each time-step, this may lead to discontinuities between successive time-steps. This issue can be addressed by incorporating two time-steps input, the onset and the end of the precipitation period, to provide more continuous predictions.

For future studies, diffusion models (Ho et al., 2020) can be explored as alternatives to our current generation-based approach. While this study demonstrates the potential of GANs for generating realistic precipitation fields, diffusion models are also known for producing sharp and detailed outputs, making them a promising alternative for future development.

Furthermore, this deep learning-based QPF model can be leveraged for additional purposes through fine-tuning, such as detecting lightning and identifying areas with heavy rainfall potential. Since the model has already learned the underlying relationship between meteorological inputs and precipitation, it can provide a strong foundation for adapting to related forecasting tasks.

*Acknowledgments*

This manuscript was prepared with the assistance of ChatGPT (OpenAI, 2025), which was used for language refinement and grammar correction. The authors take full responsibility for the content.

*Data Availability Statement.*

The code used in this study is available at https://github.com/hunter3789/Deep-Learning-QPF. The training and validation datasets used in this publication are accessible at https://osf.io/ehwmv/files/osfstorage.